\documentclass[twocolumn,prl,showpacs,preprintnumbers,amsmath,amssymb]{revtex4}

\usepackage{graphicx}
\usepackage{dcolumn}
\usepackage{bm}
\usepackage{epsfig,float,afterpage,amssymb,wrapfig,psfrag}

\newcommand{\bra}{\langle}
\newcommand{\cl}{\chi_{\mathrm{loc}}}
\newcommand{\dcl}{\chi''_{\mathrm{loc}}}
\newcommand{\down}{\downarrow}
\newcommand{\ket}{\rangle}
\newcommand{\kgc}{K_0 g_c}

\newcommand{\kgs}{(K_0 g)^2}
\newcommand{\kk}{\bm{k}}
\newcommand{\nb}{N_b}
\newcommand{\ns}{N_s}
\newcommand{\qq}{\bm{q}}
\newcommand{\ra}{\rightarrow}

\newcommand{\up}{\uparrow}
\newcommand{\w}{\omega}

\begin{document}

\preprint{}

\title{Numerical renormalization-group study of the Bose-Fermi Kondo model}

\author{Matthew T. Glossop}
\author{Kevin Ingersent}
\affiliation{Department of Physics, University of Florida,
Gainesville, Florida 32611-8440, USA}

\date{\today}
\begin{abstract}
We extend the numerical renormalization-group method to
Bose-Fermi
Kondo models (BFKMs), describing a local moment coupled to a
conduction band \textit{and} a dissipative bosonic bath.
 We apply the method to the Ising-symmetry BFKM with a bosonic bath
spectral
function $\eta(\w)\propto \w^s$, of interest in connection with
heavy-fermion
 criticality.
For $0<s<1$, an interacting critical point, characterized by hyperscaling
of exponents and
$\w/T$-scaling, describes a quantum phase transition between
Kondo-screened and
localized phases.
Connection is made to other results for the BFKM and the spin-boson model.
\end{abstract}


\maketitle

Dynamical competition between local and spatially extended degrees of freedom
provides a possible mechanism for the intriguing non-Fermi liquid behaviors
observed in cuprate high-temperature superconductors \cite{hightc} and in
heavy-fermion systems near a quantum phase transition (QPT) \cite{Stewart:01}.
The essence of this competition is embodied in a new class of quantum impurity
problems: Bose-Fermi impurity models of a local moment coupled both to a
fermionic band and a dissipative bosonic bath.

The key physics underlying heavy-fermion QPTs---interplay between screening of
$f$-shell moments by conduction electrons and magnetic ordering of those
moments due to the Ruderman-Kittel-Kasuya-Yosida interaction---is contained in
the Kondo lattice model (KLM) \cite{Coleman:99,Coleman:01,lcqpt}.
The KLM has been studied via an extended dynamical mean-field theory (EDMFT)
\cite{lcqpt,Grempel:Zhu:Sun} that maps the lattice onto a
self-consistently determined BFKM in which bosons represent the fluctuating
effective magnetic field generated by other $f$ moments.
Assuming the bosonic bath has a sub-Ohmic spectrum $\eta(\w)\propto\w^s$
$(s<1)$, a perturbative (in $\epsilon\equiv 1-s$) renormalization-group study
\cite{lcqpt} has found two types of antiferromagnetic ordering transition: a
QPT of the usual spin-density-wave type \cite{Sachdev:99}; and a
\textit{locally critical} QPT, 
characterized by $s=0^+$, at which divergence of the spatial correlation length
coincides with critical local-moment fluctuations.
There is growing evidence for the latter scenario in several materials,
particularly from neutron scattering results on CeCu$_{5.9}$Au$_{0.1}$
\cite{neutrons}.
Similar mappings to an effective Bose-Fermi impurity model have been applied to
the \textit{disordered} KLM \cite{Tanaskovic:04} and to a $t$-$J$ model
of a highly incoherent metal close to a Mott transition, of possible relevance
to the pseudogap phase of the cuprates \cite{Haule:02}. 
The BFKM also describes certain dissipative mesoscopic qubit devices, e.g.,
a noisy quantum dot, where the bosonic bath represents gate-voltage
fluctuations \cite{LeHur:04}. 

A satisfactory description of all these problems hinges on proper solution
of the BFKM.
The identification of a locally critical QPT in \cite{lcqpt} was based on
perturbative results for $\epsilon\ra 1$.
Most other techniques developed for conventional (pure-fermionic) impurity
models \cite{Hewson:93} are inapplicable or suffer from significant
limitations.
For example, nonperturbative quantum Monte Carlo studies
\cite{Grempel:Zhu:Sun} of the KLM have reached conflicting
conclusions concerning the existence of a locally critical QPT due
to inherent limitations in accessing the lowest temperatures. 

To meet the need for solutions of the BFKM that are reliable for all bath
spectra and all temperatures $T$, we have turned to Wilson's numerical
renormalization-group (NRG) technique \cite{Wilson:75}, which has
hitherto provided a controlled treatment of a range of fermionic impurity and
lattice problems \cite{Hewson:01}, and was
recently applied \cite{Bulla:03,Bulla:04} to the spin-boson model (SBM) of a
two-level system coupled to a dissipative bosonic bath \cite{Legget:87}.
Below we report a nontrivial development of the NRG that incorporates coupling
of an impurity to both a conduction band \textit{and} a bosonic bath.
We give results for the BFKM with Ising-symmetry bosonic couplings, the case
most relevant to CeCu$_{5.9}$Au$_{0.1}$ and a noisy quantum dot. For bosonic
bath exponents $0<s<1$, the model displays critical properties identical to
those of the SBM, with exponents satisfying hyperscaling relations and
$\w/T$-scaling in the dynamics.
We show directly the destruction of the Kondo resonance at the quantum critical
point (QCP). 
Importantly, this work opens the way for future decisive studies of the KLM,
obtained by imposing EDMFT self-consistency on the BFKM solution.

We study the Hamiltonian
$H=H_F+H_B+H_{\mathrm{int}}$,
 with
\begin{eqnarray}
H_F&=&\sum_{\kk,\sigma}
\varepsilon_{\kk}c^{\dagger}_{\kk\sigma}c_{\kk\sigma},
\ \ \ \ H_B=
\sum_{\qq}w_{\qq}\phi^{\dagger}_{\qq}\phi_{\qq},\nonumber\\
H_{\mathrm{int}}&=&J \bm{S} \cdot \bm{s}_c +
g S_{z}\sum_{\qq} (\phi_{\qq} + \phi^{\dagger}_{-\qq}),
\label{hamilt}
\end{eqnarray}
which describes a spin-$\frac{1}{2}$ local moment $\bm{S}$ interacting with both
the on-site spin $\bm{s}_c$ of a fermionic band, and, via $S_{z}$,
with a bosonic bath. The essential physics of the model, based on perturbative
RG and large-$N$ results \cite{Smith:99,Sengupta:00,Zhu:02,Zarand:02}, is the
competition between the Kondo coupling $J$, which tends to screen the
impurity moment, and the bosonic coupling $g$, which favors an unscreened spin.
The fermionic band is assumed to be featureless with a density of states
$\rho(\varepsilon)=\rho_0$ for $|\varepsilon|< D$.
The impurity-boson interaction is embodied entirely in
$B(\w) = \pi g^2 \eta(\w) = B_0 \ \w^s$ for $0\le \w<\w_0$, where $s > -1$.
In the EDMFT problem of future interest, $s$ will be determined self-consistently from
the form of the local spin susceptibility $\cl(\w)$ and
will in general take a non-Ohmic value ($s\ne 1$) \cite{lcqpt}.
To connect with previous work \cite{Zhu:02}, we write $B_0=\kgs$, and for
convenience we set $\w_0 = D = 1$ in the following.

\paragraph{NRG scheme.}
The band (bath) continuum of energies $|\varepsilon|<1$ ($0\le\omega<1$)
is replaced by a discrete set, $\pm\Lambda^{-n}$ ($\Lambda^{-n}$) for
$n=0,\ 1,\ 2,...$, where $\Lambda>1$ parameterizes the discretization.
Eq.~(\ref{hamilt}) may then be cast in a chain form, in close analogy to the
pure-fermionic NRG, introducing a discretization error that vanishes as
$\Lambda \ra 1$ \cite{Wilson:75}:
\begin{eqnarray}
H&=&\sum_{n,\sigma} \left[ \varepsilon_n f_{n\sigma}^{\dagger}f_{n\sigma}
+\tau_n(f_{n\sigma}^{\dagger}f_{n-1,\sigma}+f_{n-1,
\sigma}^{\dagger}f_{{n}\sigma}) \right] \nonumber \\
&+&\sum_m \left[ E_m b_m^{\dagger}b_m + T_m(b_m^{\dagger}b_{m-1}
+b^{\dagger}_{m-1}b_m) \right] \\
\nonumber
&+&\rho_0J \sum_{\sigma,\sigma'}
f_{0\sigma}^{\dagger}\sigma_{\sigma\sigma'}f_{0\sigma'}\cdot \bm{S}
\nonumber
+ K_0g \sqrt{\frac{F^2}{\pi}} S_z(b_0^{\dagger}+b_0).
\label{chain}
\end{eqnarray}
Here, $f_{0\sigma}$ ($b_0$) annihilates the unique combination of fermions
(bosons) that couples to the impurity.
The tight-binding coefficients $\varepsilon_n$, $\tau_n$, $E_m$ and $T_m$
($m,n=0,1,2,...$) encoding all information about the band and
the bath are determined via a Lanczos procedure. The decay
of $\varepsilon_n$ and $\tau_n$ as $\Lambda^{-n/2}$ for large $n$ is what allows
the iterative solution of the conventional Kondo problem via
diagonalization of progressively longer chains \cite{Wilson:75}. Since the bath
spectral function $\eta(\w)$ vanishes for $\w<0$ the coefficients $E_m$ and
$T_m$ decay faster, as $\Lambda^{-m}$ for large $m$, as discussed previously
for the SBM \cite{Bulla:03}.

We solve the discretized BFKM via an iterative scheme that adds a site to
the fermionic chain at each step, but extends the bosonic chain only at
every second step.
Since $T_{2n} \sim \tau_n$, this ensures that fermions and bosons of the same
energy scale are treated at the same step.
The number of particles per bosonic site is in principle unlimited, but is
restricted in practice to a finite maximum $\nb$.
For the conventional Kondo model, the basis must be truncated after only a few
iterations, saving only the lowest $\ns$ eigenstates of one step to
form the basis (of $4\ns$ states) for the next step.
Truncation is required even sooner in the BFKM, where the basis grows
by a factor of $4(\nb+1)$ at each step at which bosons are added.

It is not obvious \textit{a priori} that the approach outlined above should
capture the physics of the BFKM QCP \cite{alternative}, but it is validated
by the results presented below.
These results are converged with respect to the truncation parameters,
$8\le \nb\le 12$ and $500\le\ns\le 2000$ typically sufficing.
Calculated critical exponents prove insensitive to the choice of $\Lambda$
(see Table~\ref{table}), so to reduce computer time $\Lambda=9$ was used,
except where noted otherwise.

\paragraph{Critical coupling.} For $s\le 1$ the NRG flow of effective couplings
has two stable limits: a Kondo fixed point at $J=\infty$, $g=0$, describing a
phase in which conduction electrons screen the local moment and the
single-particle spectrum exhibits a Kondo resonance; and a ``localized'' fixed
point at $J=0$, $g=\infty$, where the impurity dynamics are controlled by the
coupling to the bath.
A third, unstable fixed point---the QCP at $J^*\ne 0$, $g^*\ne 0$---is reached
for bare couplings lying precisely on the boundary $g=g_c(J)$ between the
Kondo and localized phases.
For $s>1$, by contrast, the Kondo fixed point is reached for all $J>0$.
At $s=1$ the critical point is Kosterlitz-Thouless-like \cite{LeHur:04},
while for $-1 < s < 0$ it is noninteracting.
We restrict attention henceforth to the range $0<s<1$ over which the QCP
exhibits properties (shown below) characteristic of an interacting
critical point.
(For the case $s=0$ of interest in connection with local criticality,
logarithms replace power laws at the QCP.)

\begin{figure}
\begin{center}
\psfrag{xaxis}[bc][bc]{$s$}
\psfrag{yaxis}[bc][bc]{$\nu^{-1}$}
\psfrag{g}[bc][bc]{$g-g_c$}
\psfrag{x1axis}[bc][bc]{$1/\sqrt{2(1-s)}$}
\psfrag{y1axis}[bc][bc]{${\nu}$}
\epsfig{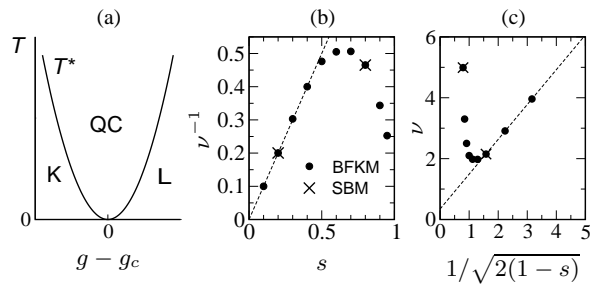}
\end{center}
\vspace{-0.5cm}
\caption{(a) Schematic phase diagram for the BFKM, showing Kondo (K),
localized (L), and quantum-critical (QC) regimes separated by a crossover scale
$T^*\propto |g-g_c|^{\nu}$ that vanishes at the QCP.
(b) and (c) Correlation length exponent $\nu$ as a function of bath exponent
$s$. Crosses (at $s=0.2$ and 0.8) show representative results for the SBM.
Dashed lines are linear fits discussed in the text.}
\label{cross}
\end{figure}

\paragraph{Correlation length exponent.} The crossover from the quantum-critical
regime to a stable low-temperature regime defines a scale that vanishes at the
critical coupling: $T^*\sim |g-g_c|^{\nu}$ [Fig.~\ref{cross}(a)].
The correlation length exponent $\nu(s)$, which can be determined from the
NRG many-body spectrum or from properties such as the local susceptibility or
the single-particle spectrum, diverges for $s\ra 0^+$ and $s\ra 1^-$,
indicative of qualitative changes at these limiting values of the bath
exponent.
For small $s$, $\nu\simeq 1/s$ [fit in Fig.~\ref{cross}(b)],
which becomes asymptotically exact as $s\ra 0$, while for $s\ra 1$ the exponent
approaches $\nu =1/\sqrt{2(1-s)} + C$ [fit in Fig.~\ref{cross}(c)].
Our $\nu$ values are in essentially exact agreement with NRG for the SBM
\cite{Bulla:03} and, for $s\ra 0$, with perturbative RG for the BFKM expanding
about $s=0$ \cite{Vojta:04}. 
\begin{figure}
\begin{center}
\epsfig{file=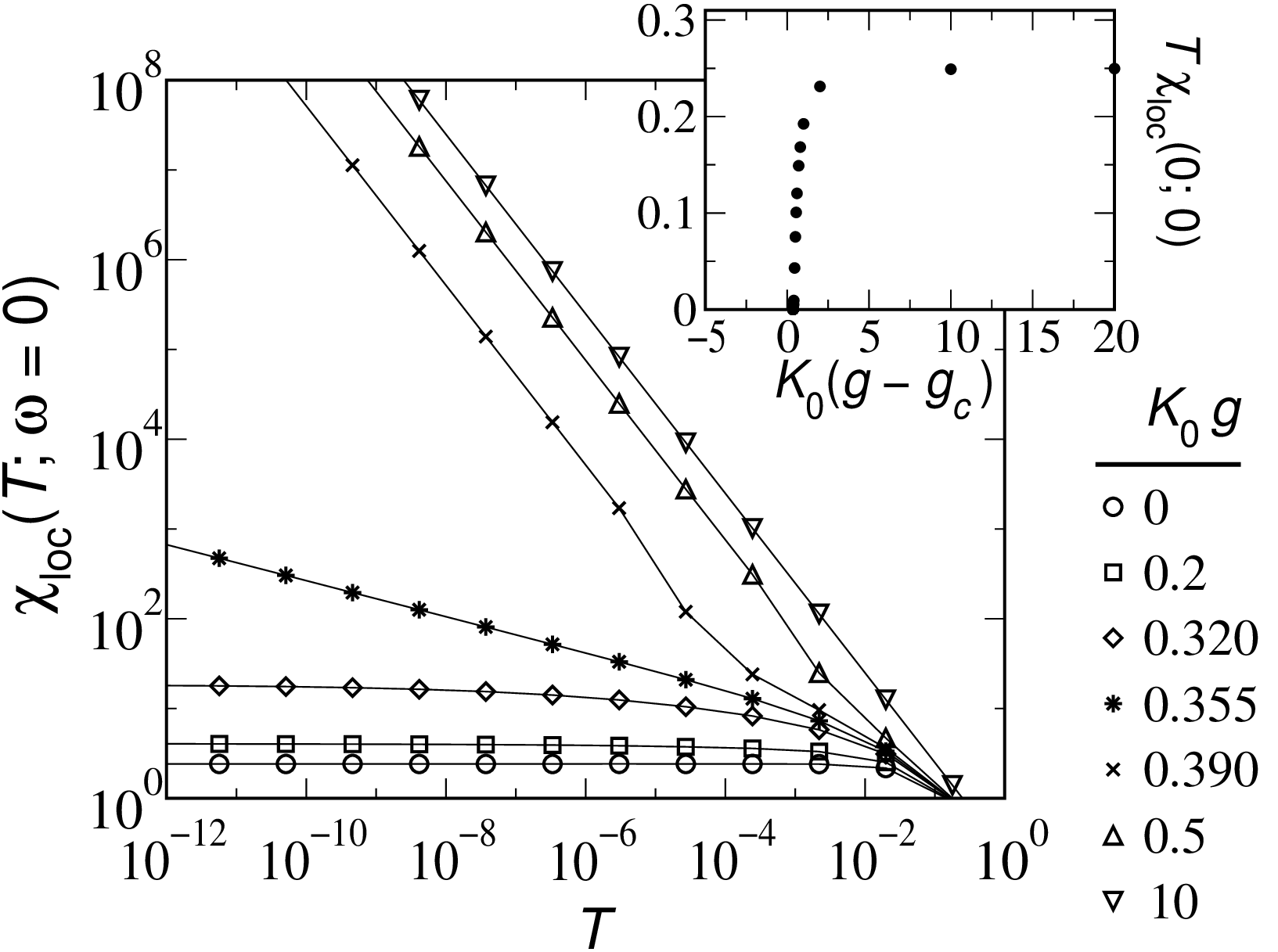, width=6cm, angle=270}
\vspace{-0.2cm}
\caption{Static local susceptibility $\cl(T; \w=0)$ vs $T$ for $s=0.2$,
and $\rho_0J=0.5$, where $\kgc\simeq 0.355$. The inset shows
the vanishing of $T\cl(0;0)$ for $g\ra g_c^+$.}
\label{static}
\end{center}
\end{figure}

\paragraph{Local magnetic response.}
Central to the EDMFT of future interest is the local susceptibility
$\cl$: the response to a magnetic field $h$ acting solely on the impurity.
Fig.~\ref{static} shows the static local susceptibility $\cl(T; \w=0)$ vs
$T$ for $s=0.2$. The two phases are readily distinguished.
For $g<g_c$, screening of the impurity moment is signaled for $T\lesssim T^*$
by a constant value $\cl(T; \w=0)\propto (g_c-g)^{-\gamma}$ that diverges as
$g\ra g_c^-$. For $g>g_c$, by contrast, there is free-moment behavior
with a Curie constant $T\cl(T; \w=0)\propto (g-g_c)^{\lambda}$ (see inset to
Fig.~\ref{static}).
At criticality, it is found that
\begin{equation}
\cl(T; \w=0, g=g_c) \propto T^{-x},
\label{x definition}
\end{equation}
with $x = s$ for all $0<s<1$, as predicted by the $\epsilon$-expansion
\cite{Zhu:02}.
We have also calculated exponents defined by
$M_{\mathrm{loc}}(g>g_c, T=0, h=0)\propto (g-g_c)^{\beta}$ and
$M_{\mathrm{loc}}(g=g_c, T=0)\propto |h|^{1/\delta}$,
where $M_{\mathrm{loc}}\equiv\bra S_z\ket$ serves as an
order parameter for the localized phase.
Table~\ref{table} shows exponents obtained for two representative cases.

Starting from a scaling ansatz for the critical part of the free energy,
$F_{\mathrm{crit}}=Tf(|g-g_c|/T^{1/\nu}, |h|/T^b)$, it is readily shown
that $\delta=(1+x)/(1-x)$, $\lambda=2\beta=\nu(1-x)$, and $\gamma = \nu x$.
The exponents obtained numerically for $0<s<1$ obey these hyperscaling
relations, indicating that the critical point is interacting \cite{Sachdev:99}.

\begin{table}[b]
\begin{ruledtabular}
\begin{tabular}{llllll}
$s$ & $\nu$ & $x$ & ${1}/{\delta}$ & $\beta$ & $\gamma$ \\
\hline
0.2 & 4.99(5) &0.200(1)& 0.665(5) & 1.99(2)& 1.00(1)\\
0.8 & 2.11(2) &0.803(3)& 0.106(2)& 0.209(1)& 1.67(2) \\
\end{tabular}
\end{ruledtabular}
\vspace*{-2ex}
\caption{\label{table}%
Critical exponents for $\Lambda=9$, $s=0.2$ and $0.8$.
Parentheses surround the estimated nonsystematic error in the last
digit.
The insensitivity to $\Lambda$ is illustrated by the $s=0.2$ values
$\nu=4.90(5)$, $x=0.200(1)$ for $\Lambda=5$, and
$\nu=4.9(1)$, $x=0.199(4)$ for $\Lambda=3$.}
\end{table}

\begin{figure}
\begin{center}
\psfrag{xaxis}[bc][bc]{ $\w/T$}
\psfrag{yaxis}[bc][bc]{$T_K^{1-s} \, T^s \, \dcl(\w,T)$}
\psfrag{x1axis}[bc][bc]{$\w$ or $T$}
\psfrag{y1axis}[bc][bc]{$\dcl(\w;T=0)$ or $\cl(T;\w=0)$}
\epsfig{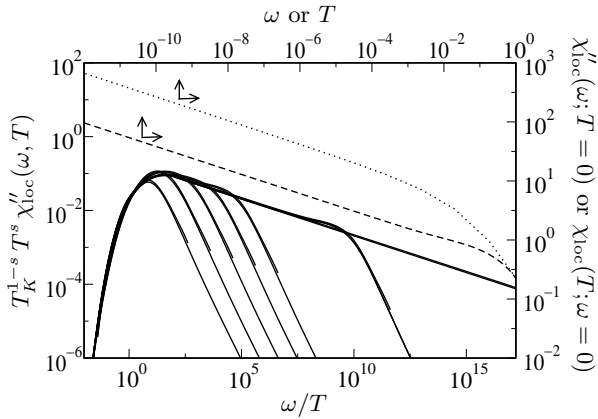}
\end{center}
\vspace{-0.5cm}
\caption{Upper-right axes: $\dcl(\w;T=0)$ (dashed) and $\cl(T;\w=0)$ (dotted) at
the QCP for $s=0.2$ and $\rho_0 J = 0.5$, showing that $x=y=s$; see
Eqs.~(\protect\ref{x definition}) and (\protect\ref{y definition}).
Lower-left axes: Finite-$T$ scaling of $\dcl(\w, T)$ at the QCP as
described
by Eq.~(\protect\ref{wovert}). Data are shown for $s=0.2$,
$\rho_0 J=0.1$ (thin line) and $0.25$ (thick line) with $g=g_c$ in each case,
and for temperatures $T/T_K = $ $10^{-1}$, $10^{-2}$, $10^{-3}$,
$10^{-4}$, $10^{-5}$, $10^{-10}$ and $10^{-20}$. The curves have a common form
for $\w/T \ll T_K/T$.}
\label{chiwovert}
\end{figure}

We now turn to the imaginary part of the dynamical local susceptibility,
$\dcl(\w)$. Figure.~\ref{chiwovert} shows $\dcl(\w; T=0, g=g_c)$ vs $\w$
for $s=0.2$.
The exponent $y$, defined via
\begin{equation}
\dcl(\w; T=0, g=g_c) \propto |\w|^{-y}\ \mbox{sgn}(\w),
\label{y definition}
\end{equation}
is found to satisfy $y=s$ for all $0<s<1$. The static local susceptibility
$\cl(T)$ is also shown, from which it is evident that $x=y$,
which likewise holds over the entire $s$ range.
Due to truncation the NRG method is known to be unreliable for $|\w|\lesssim T$
\cite{Costi:Ingersent}. However, the result
$y=x$ for $0<s<1$ (obtained in \cite{Grempel:Zhu:Sun} for the particular case
$s=0^+$) is consistent with $\w/T$-scaling at the QCP and reflects the
interacting nature of the fixed point \cite{Sachdev:99}.
Specifically, we find for $\w\ll T_K$ that
\begin{equation}
T_K\dcl(\w,T;g=g_c )\sim \left(\frac{T}{T_K}\right)^{-s}
{\phi_s}\left(\frac{\w}{T}\right) ,
\label{wovert}
\end{equation}
with $\phi_s$ a universal function of its argument as demonstrated in the lower
part of Fig.~\ref{chiwovert}. A similar result has recently been obtained for
the multi-channel BFKM in the large-$N$ limit \cite{Zhu:04}.

An important observation is that all critical exponents of the Ising-symmetry
BFKM are in numerical agreement with those of the corresponding sub-Ohmic SBM
\cite{Bulla:03,Bulla:04}. This clearly indicates that that the two QCPs belong
to the same universality class [as implied in
Ref.~\onlinecite{Grempel:Zhu:Sun}(a)].

\paragraph{Spectral function.} Finally, we calculate
$A(\w)=-\frac{1}{\pi}\mbox{Im}\bra\bra
d;d^{\dagger}\ket\ket_{\w}$
for the Ising-symmetry Bose-Fermi Anderson model described by
\begin{eqnarray}
H_{\mathrm{int}}&=&\sum_{\sigma}(\varepsilon_d+\mbox{$\frac{1}{2}$}
U n_{d,-\sigma})n_{d\sigma}
+V \sum_{\kk,\sigma}(d^{\dagger}_{\sigma}c_{\kk\sigma}+\mbox{h.c.})\nonumber\\
&+& \frac{1}{2}g(n_{d\up}-n_{d\down})\sum_{\qq}(\phi_{\qq} +
\phi^{\dagger}_{-\qq}).
\end{eqnarray}
Here, $n_d = d^{\dagger}_{\sigma} d_{\sigma}$ is the spin-$\sigma$ occupancy
of an impurity level of energy $\varepsilon_d$, $U$ is the on-site interaction,
and $V$ is the hybridization between the impurity and the conduction band.
The coupling to the bosons is again through the $z$ component of the
effective impurity spin.
In the regime $U,|\varepsilon_d|\gg\Gamma_0 \equiv\pi \rho_0 V^2$ where charge
fluctuations are suppressed, this Anderson model maps onto the BFKM under a
Schrieffer-Wolff transformation \cite{Hewson:93}.

\begin{figure}
\begin{center}
\psfrag{xaxis}[bc][bc]{$\w$}
\psfrag{yaxis}[bc][bc]{$A(\w)$}
\epsfig{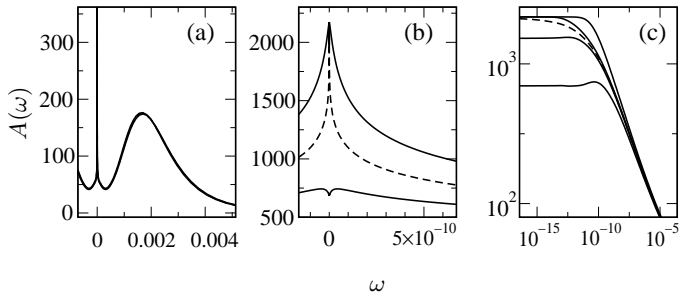}
\caption{Single-particle spectrum vs $\w$ for bath exponent $s=0.7$,
impurity parameters $U=-2\varepsilon_d=0.002$ [such that $A(\w)=A(-\w)$],
hybridization strength $\Gamma_0=1.5 \times 10^{-4}$, and discretization
$\Lambda=3$.
Shown for $B_0-B_{0,c} = 0$ (dashed line),$\ \pm 10^{-4}$ on (a) the scale of
$U$ and (b) the scale set by the central feature. (c) Shown on a logarithmic
scale for $B_0-B_{0,c}=0$ (dashed line), $\pm 10^{-5}$, $\pm 10^{-4}$.}
\label{spec}
\end{center}
\end{figure}

The QCP is manifest in the single-particle spectrum as a collapse of
the central Kondo resonance.
Fig.~\ref{spec} shows $A(\w)$ vs $\w$ close to and at the QCP.
On the scale of the Hubbard satellite bands [Fig.~\ref{spec}(a)],
the spectra are barely distinguishable.
Fig.~\ref{spec}(b) shows the same spectra on the scale of the central feature.
For $B_0 < B_{0,c}$ the integrity of the Kondo resonance is preserved; it
remains pinned at the Fermi level, $A(\w=0)=(\pi\Gamma_0)^{-1}$, as required by
Fermi-liquid theory for the conventional Anderson model \cite{Hewson:93}.
Close to the transition, the lowest-frequency (Fermi-liquid) behavior
$1-\pi\Gamma_0 A(\w)\propto \w^2$ crosses over above a scale
$|\omega|\approx\w^*$ to a power law characteristic of the QCP itself.
For $B_0=B_{0,c}$ [dashed line in Fig.~\ref{spec}(b)], $\w^*=0$, i.e., the
spectrum remains pinned but the lowest-frequency behavior is non-Fermi-liquid
in nature.
For $B_0>B_{0,c}$ the Kondo resonance collapses and $A(\w=0)$
decreases with increasing $B_0$. Near the critical coupling there is a
crossover, seen as a weak maximum in $A(\w)$ at $|\w|=\w^*$
[Fig.~\ref{spec}(c)], to the quantum-critical behavior described above.
The correlation length exponent may
also be calculated from the vanishing of $\w^*$ at the transition, and
agrees with the results of Fig.~\ref{cross} to within numerical error.

\paragraph{Summary.} We have developed the NRG method to treat coupling of
a quantum impurity to both a conduction band and a bosonic bath, as described
by Bose-Fermi Kondo and Anderson models.
Our approach provides a good account of the critical properties of the
Ising-symmetry BFKM with a power-law bath spectrum.
Over the range of bath exponents $0<s<1$, the competition between fermionic and
bosonic couplings gives rise to an interacting fixed-point, exhibiting
hyperscaling of critical exponents and $\w/T$ scaling in the impurity
dynamics, which belongs to the same universality class as the critical point
of the SBM.
The method can readily be extended to models having multiple bosonic baths,
such as the $XY$ and isotropic BFKMs.
This work also opens the way for EDMFT treatments of the Kondo
lattice down to $T=0$ that will shed much-needed light on the possibility of
local criticality in heavy-fermion systems.

We thank Q.~Si for useful comments.
This work was supported in part by NSF Grant DMR--0312939.


\begin{thebibliography}{15}
\bibitem{hightc}
M.\ R.\ Norman and C.\ P\'{e}pin,
Rep.\ Prog.\ Phys.\ \textbf{66}, 1547 (2003).

\bibitem{Stewart:01}
G.\ R.\ Stewart,
Rev.\ Mod.\ Phys.\ \textbf{73}, 797 (2001).

\bibitem{Coleman:99}
P.\ Coleman,
Physica B \textbf{259-261}, 353 (1999).

\bibitem{Coleman:01}
P.\ Coleman, C.\ P\'{e}pin, Q.\ Si, and R.\ Ramazashvili,
J.\ Phys.\ Condens.\ Matter \textbf{13}, R723 (2001).

\bibitem{lcqpt}
Q.\ Si, S.\ Rabello, K.\ Ingersent, and J.\ L.\ Smith,
Nature (London) \textbf{413}, 804 (2001);
Phys.\ Rev.\ B \textbf{68}, 115103 (2003).

\bibitem{Grempel:Zhu:Sun}
(a) D.\ R.\ Grempel and Q.\ Si,
Phys.\ Rev.\ Lett.\ \textbf{91}, 026401 (2003); 
(b) Z.-X.\ Zhu, D.\ R.\ Grempel, and Q.\ Si,
Phys.\ Rev.\ Lett.\ \textbf{91}, 156404 (2003);
(c) P.\ Sun and G.\ Kotliar,
Phys.\ Rev.\ Lett.\ \textbf{91}, 037209 (2003).

\bibitem{Sachdev:99}
S.\ Sachdev, \textit{Quantum Phase Transitions}
(Cambridge University Press, Cambridge, U.K., 1999).




\bibitem{neutrons}
H.\ von L{\"{o}}hneysen \textit{et al.},
Phys.\ Rev.\ Lett.\ \textbf{72}, 3262 (1994);
A.\ Schr\"{o}der \textit{et al.},
\textit{ibid.} \textbf{80}, 5623 (1998);
O.\ Stockert \textit{et al.},
\textit{ibid.} \textbf{80}, 5627 (1998);
A.\ Schr\"{o}der \textit{et al.},
Nature (London) \textbf{407}, 351 (2000).



\bibitem{Tanaskovic:04}
D.\ Tanaskovi{\'{c}}, V.\ Dobrosavljevi{\'{c}}, and E.\ Miranda,
preprint cond-mat/0412100.

\bibitem{Haule:02}
K.\ Haule, A.\ Rosch, J.\ Kroha and P.\ W{\"{o}}lfle,
Phys.\ Rev.\ Lett.\ \textbf{89}, 236402 (2002).

\bibitem{LeHur:04}
K.\ Le~Hur,
Phys.\ Rev.\ Lett.\ \textbf{92}, 196804 (2004).






\bibitem{Hewson:93}
A.\ C.\ Hewson, \textit{The Kondo Problem to Heavy Fermions}
(Cambridge University Press, Cambridge, U.K., 1993).






\bibitem{Wilson:75}
K.\ G.\ Wilson,
Rev.\ Mod.\ Phys.\ \textbf{47}, 773 (1975).


\bibitem{Hewson:01}
A.\ C.\ Hewson, S.\ C.\ Bradley, R.\ Bulla, and Y.\ Ono,
Int.\ J.\ Mod.\ Phys.\ B \textbf{15}, 2549 (2001).

\bibitem{Bulla:03}
R.\ Bulla, N.\ Tong, and M.\ Vojta,
Phys.\ Rev.\ Lett.\ \textbf{91}, 170601 (2003).

\bibitem{Bulla:04}
R.\ Bulla, H.\ Lee, N.\ Tong, and M.\ Vojta,
Phys.\ Rev.\ B \textbf{71}, 045122 (2005).

\bibitem{Legget:87}
A.\ J.\ Leggett \textit{et al.},
Rev.\ Mod.\ Phys.\ \textbf{59}, 1 (1987).


\bibitem{Smith:99}
J.\ L.\ Smith and Q.\ Si,
Europhys.\ Lett.\ \textbf{45}, 228 (1999).

\bibitem{Sengupta:00}
A.\ M.\ Sengupta,
Phys.\ Rev.\ B \textbf{61}, 4041 (2000).

\bibitem{Zhu:02}
L.\ Zhu and Q.\ Si,
Phys.\ Rev.\ B \textbf{66}, 024426 (2002).

\bibitem{Zarand:02}
G.\ Zar\'{a}nd and E.\ Demler,
Phys.\ Rev.\ B \textbf{66}, 024427 (2002).

\bibitem[{alt()}]{alternative}
\bibinfo{note}{The chain mapping of the bosons may fail deep inside the BFKM
localized regime due to a divergence of the mean site occupancy.
In this case, an alternative mapping proposed for the SBM
\protect\cite{Bulla:04}} may prove more successful.

\bibitem{Vojta:04}
M.\ Vojta, N.\ Tong, and R.\ Bulla, 
Phys.\ Rev.\ Lett.\ \textbf{94}, 070604 (2005).

\bibitem{Costi:Ingersent}
T.\ A.\ Costi, A.\ C.\ Hewson, and V.\ Zlati\'{c},
J.\ Phys.\ Condens.\ Matter \textbf{6}, 2519 (1994); K.\ Ingersent and Q.\ Si,
Phys.\ Rev.\ Lett.\ \textbf{89}, 076403 (2002).

\bibitem{Zhu:04}
L.\ Zhu, S.\ Kirchner, Q.\ Si, and A.\ Georges,
Phys.\ Rev.\ Lett.\ \textbf{93}, 267201 (2004).



\end{thebibliography}

\end{document}